\documentclass{Interspeech2024}

\usepackage{multirow}
\usepackage{stfloats}
\usepackage{float}
\usepackage{amsmath,graphicx,booktabs,amsfonts,amssymb,utfsym,multirow,makecell,url,color, colortbl,subcaption,siunitx}
\usepackage{soul}

\interspeechcameraready

\title{Towards Supervised Performance on Speaker Verification with Self-Supervised Learning by Leveraging Large-Scale ASR Models}

\name{Victor}{Miara}
\name{Theo}{Lepage}
\name{Reda}{Dehak}

\address{
  EPITA Research Laboratory (LRE), France
}
\email{victor.miara@epita.fr, theo.lepage@epita.fr, reda.dehak@epita.fr}


\begin{document}

\maketitle

\begin{abstract}
Recent advancements in Self-Supervised Learning (SSL) have shown promising results in Speaker Verification (SV). However, narrowing the performance gap with supervised systems remains an ongoing challenge. Several studies have observed that speech representations from large-scale ASR models contain valuable speaker information. This work explores the limitations of fine-tuning these models for SV using an SSL contrastive objective in an end-to-end approach. Then, we propose a framework to learn speaker representations in an SSL context by fine-tuning a pre-trained WavLM with a supervised loss using pseudo-labels. Initial pseudo-labels are derived from an SSL DINO-based model and are iteratively refined by clustering the model embeddings. Our method achieves 0.99\% EER on VoxCeleb1-O, establishing the new state-of-the-art on self-supervised SV. As this performance is close to our supervised baseline of 0.94\% EER, this contribution is a step towards supervised performance on SV with SSL.
\end{abstract}

\newcommand\blfootnote[1]{%
  \begingroup
  \renewcommand\thefootnote{}\footnote{#1}%
  \addtocounter{footnote}{-1}%
  \endgroup
}

\blfootnote{The code associated with this article is publicly available at\\ \url{https://github.com/theolepage/wavlm_ssl_sv}.}

\section{Introduction}

Speaker Recognition (SR) is the process of recognizing the identity of the person speaking in an audio speech utterance. A typical application is Speaker Verification (SV), which aims to determine whether two audio utterances originate from the same speaker. The objective of SR systems is to learn speaker representations by maximizing inter-speaker distances while minimizing intra-speaker variances and being robust to extrinsic variabilities. With the emergence of deep learning, conventional machine learning methods such as i-vectors \cite{dehak2011IVector} have been outperformed by DNN models \cite{chungVoxCeleb2DeepSpeaker2018, chungDefenceMetricLearning2020} such as the d-vectors \cite{varianiDeepNeuralNetworks2014}, the x-vectors \cite{snyder2018XVectors}, and more recently ECAPA-TDNN \cite{desplanques2020ECAPATDNN}. These methods are trained to match an input utterance to the corresponding speaker identity in a supervised manner. However, this training scheme represents a notable constraint due to the scarcity and expense of getting annotated training samples determining the performance of DNN models.

Self-Supervised Learning (SSL) methods have emerged as a solution to this problem by learning meaningful representation directly from the input data. SSL has been shown to enhance the scalability potential of models by leveraging the abundance of unlabeled data available. When applied to SV, most frameworks rely on contrastive learning, such as AP+AAT \cite{huh2020AATUnsupervisedSR}, SimCLR \cite{zhang2021SSLSVSimCLR} and MoCo \cite{xia2021SSLSVMoCo}. They assume that two utterances, from distinct audio files sampled randomly in the training set do not belong to the same speaker. Based on this assumption, they aim to maximize the similarity in the embedding space between positive pairs and minimize the similarity between negative pairs (constructed by sampling utterances from different audio files). Other methods have been explored and successfully applied to SV \cite{ravanelli2019LearningSpeakerRepresentationsMI, lepage2022LabelEfficientSelfSupervisedSV, mohammadamini2022BarlowTwinsSSLSV}, notably techniques relying on knowledge distillation such as DINO which represents the state-of-the-art \cite{chen2023RDINO, cho2022jesus, zhang2022C3DINO}. DINO is based on a self-distillation training process that iteratively maximizes the cross-entropy between the output of a student and a teacher branch on multiple different views of the same utterance. All these methods rely on extensive data augmentation to avoid encoding channel characteristics as positive segments are extracted from the same audio files.

The field of Automatic Speech Recognition (ASR) has witnessed significant advancements by pre-training large Transformer-based models in a self-supervised way on speech prediction with masked units, inspired by NLP methods \cite{devlin2018BERT}. wav2vec \cite{schneider2019wav2vec, baevski2020wav2vec2.0}, HuBERT \cite{hsu2021HuBERT}, and WavLM \cite{chen2022WavLM} have been major milestones in ASR. Training these large models from scratch requires immense computational capacity. Nonetheless, due to their massive number of trainable parameters and extensive pre-training, they have been shown to generalize well to SR when fine-tuned \cite{chen2022SSLASRBenefitSR, chowdhury2024WhatDoASRLearnAboutSR, fan2021ExploringWav2vec2.0SV}. WavLM stands out as its training objective compels the model to encode information relevant to other speech-processing tasks, enabling it to achieve state-of-the-art in SV when fine-tuned with an ECAPA-TDNN. WavLM MHFA \cite{peng2023MHFA} replaces the ECAPA-TDNN with a lightweight attention-based back-end while reducing the number of parameters. Therefore, we observe a paradigm shift from relying on task-specific models trained on small datasets to harnessing speech representations for SR.



We propose a strategy to learn speaker representations by fine-tuning a pre-trained WavLM with the MHFA back-end using a self-supervised strategy. Because of the impracticability of using an SSL contrastive objective in an end-to-end fashion, we generate pseudo-labels from DINO to fine-tune the ASR model with a supervised loss. Compared to \cite{chen2022WavLM, peng2023MHFA}, our method is fully self-supervised as our fine-tuning does not rely on speaker identity labels from the train set. While other equivalent SSL techniques \cite{han2022DLG-LC, cho2021JHU} use standard SR models, we use an ASR model as the backbone. Thus, this contribution is the first to apply SSL to a large-scale pre-trained ASR model for SV. Our approach achieves 0.99\% EER on VoxCeleb1-O, representing the new state-of-the-art on this task with SSL, and narrows the gap with supervised performance.

Our complete framework is described in Section~\ref{sec:methods_pseudolabels}. Section~\ref{sec:ExperimentalSetup} describes our experimental setup. The results of our system on SV are presented and compared to other methods from the literature in Section~\ref{sec:ResultsAndDiscussions}. Section~\ref{sec:Conclusions} concludes the article.
\begin{figure*}
  \centering
  \includegraphics[width=\textwidth]{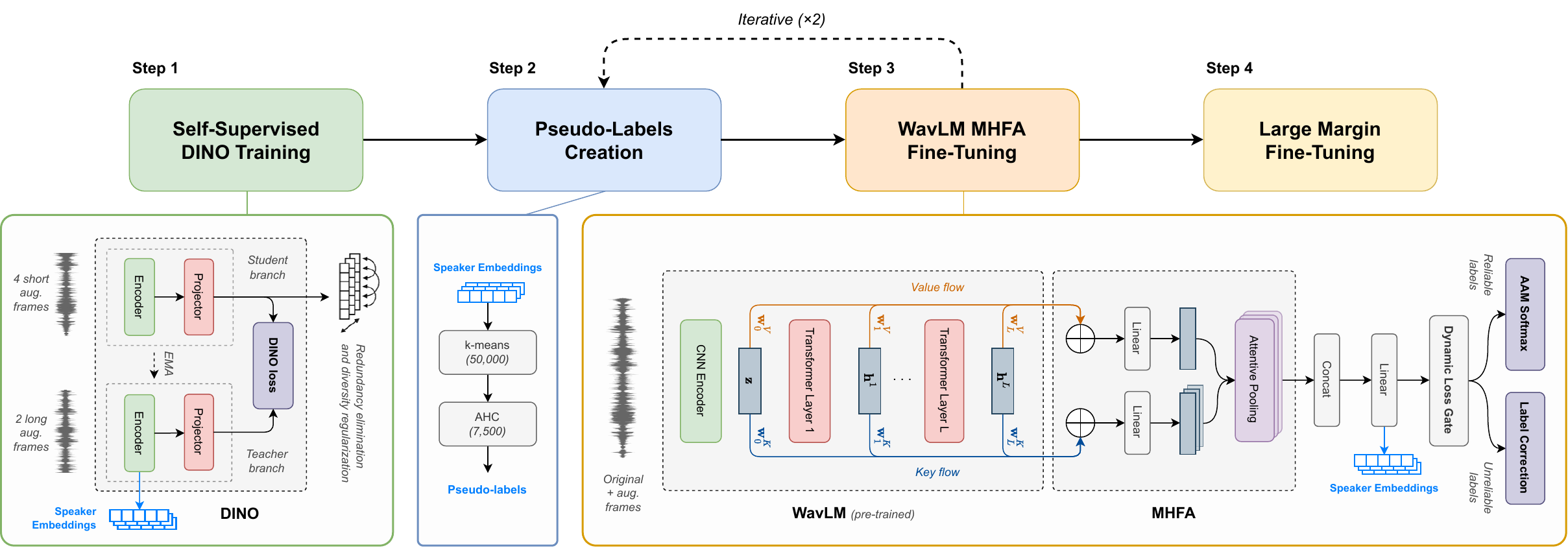}
  \caption{Our complete training framework to leverage speech representations for SV using a self-supervised strategy.}
  \label{fig:framework}
\end{figure*}

\section{Methods}

\subsection{Self-supervised learning for speaker verification}
\label{sec:methods_ssl}

\subsubsection{Contrastive learning (SimCLR)}

SSL contrastive learning assumes that two randomly sampled utterances belong to different speakers. The learning task aims at maximizing the similarity within positive pairs, derived from the same utterances, while maximizing the distance between negative pairs, sampled from the mini-batch. Following, SimCLR \cite{chen2020SimCLR}, we rely on a siamese architecture to produce a pair of embeddings for a given unlabeled utterance and apply extensive data-augmentations which is fundamental to produce robust representations. Let $i \in I \equiv\{1 \ldots 2 N\}$ be the index of an utterance from the mini-batch of size $N$, $j(i)$ be the index of the other augmented frame from the same audio and $A(i) \equiv I \setminus \{i\}$. \textit{NT-Xent} loss is defined as
\begin{equation}
    \mathcal{L}_{\text {NT-Xent}}=- \frac{1}{2N} \sum_{i \in I} \log \frac{\ell \left(\boldsymbol{z}_i, \boldsymbol{z}_{j(i)}\right)}{\sum\limits_{a \in A(i)} \ell \left(\boldsymbol{z}_i, \boldsymbol{z}_a \right)},
\end{equation}
where $\boldsymbol{z}_i$ is an embedding and $\ell(\boldsymbol{u}, \boldsymbol{v})=e^{\cos\left(\theta_{\boldsymbol{u}, \boldsymbol{v}}\right) / \tau}$.


\subsubsection{Knowledge distillation (DINO)}




DINO \cite{caron2021DINO} is a self-distillation framework where a student network is trained to match the output of a teacher network. From an utterance $x$, a set of different views $V$ are generated with data-augmentation resulting in four \textit{local} segments and two \textit{global} segments ($x_1^g$ and $x_2^g$). All views are fed through the student branch, while only global views are fed through the teacher network to encourage ”local-to-global” correspondences. Both networks share the same architecture, composed of an encoder, where speaker embeddings are extracted, and a projector. The student and teacher output a $K$ dimensional feature ($P_s$ and $P_t$, respectively) that is normalized with a temperature softmax over the feature dimension. The objective function minimizes the cross-entropy between the two probability distributions and is defined as
\begin{equation}
    \mathcal{L}_{\text {DINO}}= \sum_{x \in\left\{x_1^g, x_2^g\right\}} \sum_{\substack{x^{\prime} \in V \\ x^{\prime} \neq x}} H\left(P_t(x), P_s\left(x^{\prime}\right)\right),
\end{equation}
where $H\left(a, b\right)=-a \log b$. During training, the teacher network is updated with an Exponential Moving Average (EMA) of the student weights. To avoid collapse (i.e. trivial solutions), \textit{sharpening} and \textit{centering} are applied to the teacher outputs. Following RDINO \cite{chen2023RDINO}, we implement redundancy elimination and diversity regularizations.

\subsection{WavLM-based speaker recognition}
\label{sec:methods_wavlm}


WavLM \cite{chen2022WavLM} is a Transformer-based ASR model pre-trained in a self-supervised way that also encodes non-ASR information such as speaker identity. During the pre-training phase, a multi-layer convolutional feature encoder takes as input the sequence $\left\{\mathbf{x}_t\right\}_{t=1}^T$ of $T$ time windows from a raw audio sample to produce $\left\{\mathbf{z}_t\right\}_{t=1}^T$. These representations are simulated noisy/overlapped with masks and fed into the Transformer encoder which outputs a series of hidden states $\left\{\mathbf{h}^l\right\}_{l=1}^L$, with $L$ the number of Transformer layers. Furthermore, the model uses gated relative position bias, enhancing its ability to focus on relevant speech characteristics. WavLM is trained on a masked speech denoising and prediction task which implicitly models speaker-related information as the objective is to predict the pseudo-labels of the original speech on masked regions. The pseudo-labels are generated by first clustering the Mel-Frequency Cepstral Coefficients (MFCCs) of the training data, and then the latent representations.

\subsubsection{WavLM with Multi-Head Factorized Attention}


Multi-Head Factorized Attention (MHFA) \cite{peng2023MHFA} back-end consists of aggregating layer-wise outputs from WavLM's transformer layers into an attentive pooling mechanism. This method relies on two components: (1) L2 regularization between the current weights and the initial pre-trained weights, preventing overfitting due to a large number of parameters; (2) layer-wise learning rate decay, following \cite{sun2020finetuneBERT}. As there is a progressive abstraction of information throughout the Transformer layers \cite{chen2022WavLM}, this latter technique allows more flexible weight modifications in higher layers to refine ASR capabilities, while ensuring lower layers preserve speaker-related information.


\subsection{Unfeasibility of end-to-end self-supervised fine-tuning}
\label{sec:methods_limitations}

\begin{figure}[t]
    \centering
    \begin{subfigure}[t]{0.49\columnwidth}
        \centering
        \includegraphics[width=\linewidth]{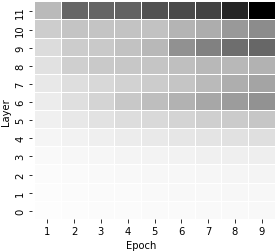}
        \caption{Supervised training}
        \label{fig:layer_change_sup}
     \end{subfigure}
     \hfill
     \begin{subfigure}[t]{0.49\columnwidth}
         \centering
         \includegraphics[width=\linewidth]{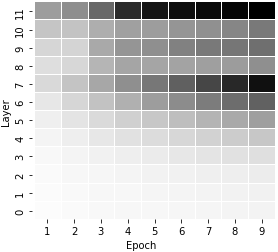}
         \caption{SSL contrastive training}
         \label{fig:layer_change_ssl}
     \end{subfigure}
    \caption{$l_2$ distance between the pre-trained and fine-tuned weights of the WavLM at different layers and epochs.}
    \label{fig:layer_change}
\end{figure}

Our experiments showed that fine-tuning WavLM directly with a self-supervised contrastive loss leads to a sub-optimal solution. Using the DINO framework was not considered due to memory limitations. Figure~\ref{fig:layer_change} reveals weight changes in Transformer layers during supervised (\textit{AAM-Softmax} loss) and self-supervised (\textit{NT-Xent} loss) training per epoch. We notice that optimizing an SSL contrastive loss leads to greater changes, especially in early layers which contain more speaker information. Considering that SSL contrastive methods are prone to learn channel-related information, as positive pairs are sampled from the same utterances, we assume that these weight changes reflect a focus on features extrinsic to speaker identity. This misalignment between the SSL contrastive loss and the downstream task, inherent to the self-supervision and amplified by WavLM learning capabilities, leads to poorer performance.



\subsection{Self-supervised fine-tuning by leveraging pseudo-labels}
\label{sec:methods_pseudolabels}

Based on this impracticability, we opt to fine-tune WavLM MHFA with pseudo-labels, initially extracted from a self-supervised model, instead of training the whole framework in an end-to-end self-supervised way. Our training framework is depicted in Figure~\ref{fig:framework} and detailed in this section.

\subsubsection{Fine-tuning WavLM MHFA with refined pseudo-labels}
After the DINO-based model training (\textbf{Step 1}), initial pseudo-labels are generated by applying a clustering algorithm on training samples' embeddings (\textbf{Step 2}). We use k-means with 50,000 clusters followed by Agglomerative Hierarchical Clustering (AHC) with 7,500 clusters, similar to \cite{cho2022jesus}. Each cluster is treated as a distinct speaker class, resulting in a list of sample-label pairs used to fine-tune WavLM MHFA (\textbf{Step 3}). Then, we iteratively generate refined pseudo-labels from the WavLM MHFA of the last iteration (\textbf{Step 2} and \textbf{Step 3}). This method allows training our framework in a self-supervised way, without ever needing the original training labels. After two iterations, we increase the input audio length and the AAM-Softmax margin to perform the Large Margin Fine-Tuning (LMFT) (\textbf{Step 4}).

\subsubsection{Handling unreliable pseudo-labels}
Similarly to \cite{tao2022LGL}, samples with incorrect pseudo-labels typically have higher loss values after a few epochs during the training. This can be attributed to the model's fast adaptation and generalization to samples with reliable pseudo-labels, representing the majority of the training data. Therefore, we apply Dynamic Loss-Gate (DLG), which ignores unreliable samples having a loss higher than a defined threshold during the training. Following \cite{han2022DLG-LC}, we implement Label Correction (LC) to avoid discarding all unreliable samples. This involves maximizing the similarity between the output distributions of clean and augmented versions of utterances, with a cross-entropy loss.
\section{Experimental setup}
\label{sec:ExperimentalSetup}
\subsection{Datasets}



The fine-tuning of the WavLM MHFA is performed on the VoxCeleb2 \cite{chung2018Vox2} \textit{dev} set, which contains 1,092,009 utterances from 5,994 speakers. We do not rely on the original speaker identity labels at any stage of the training. We apply online data augmentation with the MUSAN \cite{snyder2015musan} noise and the Room Impulse Response and Noise Database \cite{ko2017rir} reverberation corpora, similar to \cite{lepage2022LabelEfficientSelfSupervisedSV}. All systems are evaluated on VoxCeleb1 \cite{nagrani2017vox1} test sets comprising 4,874 utterances from 40 speakers.

\subsection{Implementation details}

\subsubsection{DINO}
DINO is trained from scratch in a self-supervised way, using the setup described in \cite{chen2023RDINO}. The encoder is an ECAPA-TDNN \cite{desplanques2020ECAPATDNN} which outputs a 512-dimensional embedding.

\subsubsection{WavLM MHFA}
The pre-trained WavLM base+\footnote{\url{https://github.com/microsoft/unilm}} \cite{chen2022WavLM} and the MHFA back-end \cite{peng2023MHFA} have respectively 94M and 2M learnable parameters. WavLM base+ comprises a CNN encoder, acting as a feature extractor, and 12 Transformer layers. The dimension of the Transformer's output is $768$, and the dimension of the MHFA back-end's output, used for speaker verification, is $256$. WavLM MHFA is trained on 2 $\times$ NVIDIA A100 80 GB GPUs for 15 epochs. The mini-batch size is set to $120$, the optimizer is Adam, and the loss is AAM-Softmax with a margin of $0.2$ and a scale of $30$. The learning rate is decreased by $5\%$ at each epoch. The input raw waveform duration is \SI{3}{\second}. Dynamic Loss-Gate (DLG) is not applied during the first 5 epochs. Onwards, the losses for each sample are recorded to dynamically compute the loss gate value $\tau_{1}$. As described in \cite{han2022DLG-LC}, a Gaussian Mixture Model (GMM) with two components is used to fit the loss distribution, and the threshold $\tau_{1}$ is set at the intersection of the two distributions. Label Correction (LC) is applied solely to unreliable samples, which have a loss value higher than $\tau_1$ and at least one output class probability greater than $\tau_2$. LC is activated 3 epochs after enabling DLG, allowing the loss distribution to have two distinct components, reliable and unreliable samples. Similar to \cite{han2022DLG-LC}, the threshold $\tau_{2}$ is set to $0.5$ and the sharpen parameter $\epsilon_{c}$ to $0.1$. For the Large Margin Fine-Tuning (LMFT), we continue the training for 2 epochs, and we set the input audio duration to \SI{5}{\second} and the margin to $0.5$.

\subsection{Evaluation protocol}
To perform SV, we extract embeddings from $15$ evenly spaced frames of \SI{3}{\second} for each utterance. The trial score is the average of the cosine similarity for all combinations of $l_2$-normalized embeddings pairs. We report the performance in terms of Equal Error Rate (EER) and minimum Detection Cost Function (minDCF) with $P_{target} = 0.01$, $C_{\text{miss}}=1$ and $C_{\text{fa}}=1$.
\definecolor{Gray}{gray}{0.9}
\setcounter{table}{2}
\begin{table*}[t]
  \caption{Evaluation of different self-supervised SV methods on VoxCeleb1 trials (Original, Extended, Hard). The results for the top rows are drawn from the literature. "\textbf{Ours}" represents the system described in this article (Section~\ref{sec:methods_pseudolabels}). "Supervised" is our baseline which corresponds to our reproduction of WavLM MHFA \cite{peng2023MHFA}. The number of iterations includes the first pseudo-labeling from DINO.}
  \label{tab:comparison_methods}
  \centering
  \begin{tabular}{lcccccccc}
    \toprule    
    \multirow{2}{*}{\textbf{Method}} & \multirow{2}{*}{\textbf{\# of iterations}} & \multicolumn{2}{c}{\textbf{VoxCeleb1-O}} & \multicolumn{2}{c}{\textbf{VoxCeleb1-E}} & \multicolumn{2}{c}{\textbf{VoxCeleb1-H}} \\
    \cmidrule(r){3-4} \cmidrule(r){5-6} \cmidrule(r){7-8}
    & & \textbf{EER (\%)} & \textbf{$\text{minDCF}_\text{0.01}$} & \textbf{EER (\%)} & \textbf{$\text{minDCF}_\text{0.01}$} & \textbf{EER (\%)} & \textbf{$\text{minDCF}_\text{0.01}$} \\
    \midrule
    \rowcolor{Gray} JHU \cite{cho2021JHU} & 4 & 1.89 & - & - & - & - & - \\
    \rowcolor{Gray} DKU \cite{cai2021DKU} & 4 & 1.81 & - & - & - & - & - \\
    \rowcolor{Gray} SNU \cite{mun2021SNU} & 4 & 1.66 & - & - & - & - & - \\
    \rowcolor{Gray} LGL \cite{tao2022LGL} & 5 & 1.66 & - & 2.18 & - & 3.76 & - \\
    \rowcolor{Gray} DLG-LC \cite{han2022DLG-LC} & 5 & 1.47 & - & 1.78 & - & 3.19 & - \\
    \midrule
    \textbf{Ours} & \textbf{3} & \textbf{0.99} & \textbf{0.0905} & \textbf{1.21} & \textbf{0.1263} & \textbf{2.35} & \textbf{0.2214} \\
    \midrule
    Supervised & - & 0.94 & 0.1179 & 0.93 & 0.1066 & 1.94 & 0.1919 \\
    \bottomrule
  \end{tabular}
\end{table*}
\setcounter{table}{0}
\begin{table}[ht]
  \caption{Fine-tuning WavLM with an SSL contrastive loss. Positive pairs are either sampled from the same audio files (SimCLR) or distinct audio files (supervised). EER and minDCF are evaluated on SV (VoxCeleb1-O).}
  \label{tab:wavlm_ssl}
  \centering
  \begin{tabular}{lcc}
    \toprule    
    \textbf{Positive pairs sampling} & \textbf{EER (\%)} & \textbf{$\text{minDCF}_\text{0.01}$} \\
    \midrule
    Same audio files (SimCLR) & 15.13 & 0.9586 \\
    Different audio files & 10.81 & 0.9377 \\
    \bottomrule
  \end{tabular}
\end{table}

\newcommand{\printN}[1]{%
    \ifnum#1>0
        \hspace{\parindent} \hspace{\parindent} \hspace{\parindent} \hspace{\parindent} %
        \printN{\numexpr#1-1\relax}%
    \fi
}

\section{Results and discussions}
\label{sec:ResultsAndDiscussions}

\subsection{Fine-tuning WavLM with an SSL contrastive loss}

In Table~\ref{tab:wavlm_ssl}, we present the results obtained on SV when fine-tuning WavLM with an SSL contrastive loss via the SimCLR framework (\textit{NT-Xent} loss). This experiment used an MLP on top of the WavLM and trained the system for 50 epochs on VoxCeleb1. Sampling positive pairs from the same utterances, following contrastive SSL frameworks, results in an EER that is not on par with equivalent methods. However, using labels to sample positive pairs from different utterances improves downstream performance significantly, which reinforces our assumption that the strong reliance of contrastive SSL on channel characteristics prevents the WavLM from focusing on speaker information during fine-tuning, as discussed in Section \ref{sec:methods_limitations}.

\subsection{Effect of the different components of our framework}

In Table~\ref{tab:ablation_study}, we report the incremental improvement of the different components of our framework on speaker verification (VoxCeleb1-O). Training the WavLM MHFA on DINO's pseudo-labels results in a 52.5\% relative reduction of the EER. This shows the benefit of leveraging speech representations by fine-tuning the ASR model with a supervised objective on the pseudo-labels obtained in a self-supervised way. The use of MHFA is motivated by the fact that speaker information is not uniformly distributed among the Transformer layers. In addition, MHFA lowers training time by limiting the number of trainable parameters compared to other back-ends. Dynamic Loss-Gate (DLG) and Label Correction (LC) improve the downstream performance to 1.22\% EER, showing the importance of handling unreliable pseudo-labels. The Iterative Clustering (IC), generating refined pseudo-labels, allows reaching 1.01\% EER after the second iteration. We opt to discard the third iteration as it results in equivalent performance. Finally, we perform the Large Margin Fine-Tuning (LMFT) step and achieve 0.99\% EER. To assess the quality of our pseudo-labels, we rely on the Adjusted Rand Index (ARI) and the Normalized Mutual Information (NMI). Initial pseudo-labels (DINO) results in 0.81 ARI and 0.95 NMI while pseudo-labels obtained from our final system results in 0.90 ARI and 0.98 NMI. This improvement proves the efficiency of our system.

\begin{table}[t]
  \caption{Incremental study of the components of our framework. Grayed rows are discarded. LMFT represents our final model. EER and minDCF are evaluated on SV (VoxCeleb1-O).}
  \label{tab:ablation_study}
  \centering
  \begin{tabular}{lcc}
    \toprule    
    \textbf{Method} & \textbf{EER (\%)} & \textbf{$\text{minDCF}_\text{0.01}$} \\
    \midrule
    DINO & 3.16 & 0.2233 \\
    \printN{1} + WavLM MHFA & 1.50 & 0.1378 \\
    \printN{2} + DLG & 1.27 & 0.1401 \\
    \printN{3} + LC & 1.22 & 0.1531 \\
    \printN{4} + IC (iter 1) & 1.17 & 0.1351 \\
    \printN{5} + IC (iter 2) & 1.01 & 0.1399 \\
    \printN{6} {\color{gray}+ IC (iter 3)} & {\color{gray}1.08} & {\color{gray}0.1340} \\
    \printN{6} + \textbf{LMFT} & \textbf{0.99} & \textbf{0.0905} \\
    \bottomrule
  \end{tabular}
\end{table}

\subsection{Comparison to other speaker verification methods}


An evaluation of different self-supervised SV methods, including our system, is presented in Table~\ref{tab:comparison_methods}. The final performance of our self-supervised framework is 0.99\% EER and 0.0640 minDCF on VoxCeleb1-O. Additionally, we achieve 1.21\% EER and 2.35\% EER on VoxCeleb1 \textit{Extended} and \textit{Hard} test sets, respectively. This result almost reaches the supervised baseline set at 0.94\% EER and 0.0655 minDCF on VoxCeleb1-O. Thus, our system significantly narrows the gap between supervised and self-supervised performance. Furthermore, we achieve state-of-the-art performance on self-supervised SV as we outperform other equivalent methods from the literature \cite{cho2021JHU, cai2021DKU, mun2021SNU, tao2022LGL, han2022DLG-LC} on all VoxCeleb1 trials. This shows the robustness of our system as we only rely on 3 iterations of iterative clustering and do not apply model fusion or calibration.
\section{Conclusions}
\label{sec:Conclusions}

In this article, we propose a fully self-supervised framework to leverage speech representations for SV. We show that fine-tuning WavLM with an SSL contrastive objective function does not capture valuable speaker information. Thus, we rely on pseudo-labels generated through self-supervision to fine-tune the pre-trained ASR model with the MHFA back-end. Our experiments on the VoxCeleb1 test set show that our method achieves state-of-the-art performance on self-supervised SV with a result close to the supervised baseline. This contribution is a step towards supervised performance with self-supervised learning on this task.
\section{Acknowledgements}

This work was performed using HPC resources from GENCI-IDRIS (Grant 2023-AD011014623) and has been partially funded by the French National Research Agency (project APATE - ANR-22-CE39-0016-05).

\bibliographystyle{IEEEtran}
\bibliography{mybib}

\end{document}